\documentclass[aps,prl,twocolumn,showpacs,preprintnumbers,amsmath,amssymb]{revtex4}
\usepackage{graphicx}% Include figure files
\usepackage{bm}% bold math

\begin{document}

\title{A Model for Superconductivity in Ferromagnetic ZrZn$_2$}
\author{M. B. Walker}
\author{K. V. Samokhin}
\affiliation{Department of Physics,University of Toronto, Toronto,
Ont. M5S 1A7 }
\date{\today}

\begin{abstract}
This article proposes that superconductivity in the ferromagnetic
state of ZrZn$_2$ is stabilized by an exchange-type interaction
between the magnetic moments of triplet-state Cooper pairs and the
ferromagnetic magnetization density. This explains why
superconductivity occurs in the ferromagnetic state only, and why
it persists deep into the ferromagnetic state. The model of this
article also yields a particular order parameter symmetry, which
is a prediction that can be checked experimentally.
\end{abstract}


\maketitle

Recently, superconducting states have been found to coexist with
ferromagnetism in the materials UGe$_2$ \cite{sax00,hux01,tat01},
ZrZn$_2$ \protect\cite{pfl01}, and URhGe \protect\cite{aok01}. The
initial discovery in UGe$_2$ was motivated by the idea that
parallel-spin (and hence spin-triplet-state) Cooper pairs would be
favored in a metallic state close to the border of ferromagnetism.
The proximity of a ferromagnetic state would give rise to
relatively strong ferromagnetic fluctuations which would promote
spin-triplet pairing.

\begin{figure}
\includegraphics[width = 2.5in, height = 2.0in]{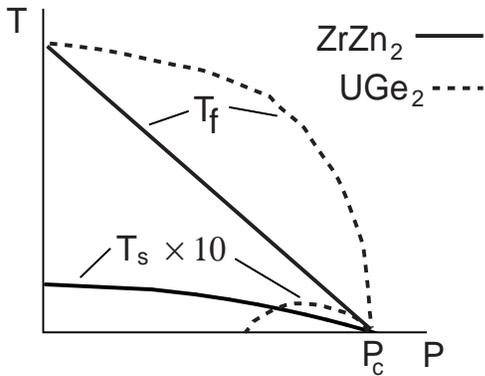}
\caption{ \label{pd} Phase diagram showing the ferromagnetic ($T_f$) and
superconducting ($T_s$) transition temperatures in ZrZn$_2$  as
functions of pressure, as derived from the model of this article
and as determined by experiment.\protect\cite{pfl01}  For clarity,
the temperature scale for the superconducting phase transition has
been multiplied by a factor of approximately 10 relative to that
for the ferromagnetic phase transition as in Ref.\
\protect\onlinecite{pfl01}.  Note that the qualitative behaviors
of both $T_f$ and $T_s$ for UGe$_2$ (sketch of data from Ref.\
\protect\onlinecite{sax00}) are quite different from those for
ZrZn$_2$.}
\end{figure}

\begin{figure}
\includegraphics{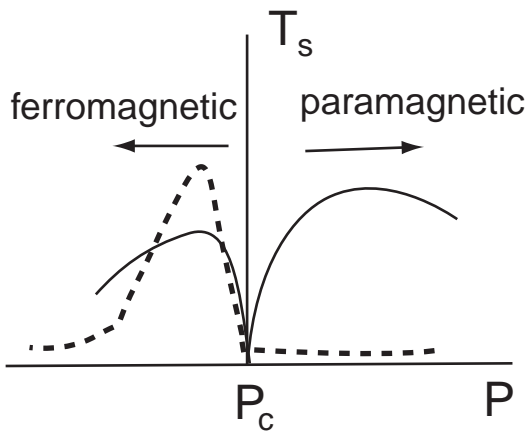}
\caption{\label{pd2}Schematic reproduction of the results of Ref.\
\protect\onlinecite{fay80} (solid line) and Ref.\
\protect\onlinecite{kir01} (dashed line) showing  theoretical
calculations of the $p$-wave superconducting transition
temperature  versus pressure P.  (Here P represents any parameter
characterizing the distance from the quantum critical point.)  The
result  of Ref.\ \protect\onlinecite{fay80} and others (see text)
were responsible for the idea that superconductivity in a
ferromagnetic state would also be accompanied by superconductivity
in the  neighboring paramagnetic state.  The very recent result of
Ref.\ \protect\onlinecite{kir01} is qualitatively similar to the
phase diagram determined for UGe$_2$ (which has a small pocket of
superconductivity close to $P_c$) but not to that for ZrZn$_2$
(where superconductivity occurs at all $P$ between $P = 0$ and $P
= P_c$).}
\end{figure}

A sketch of the phase diagram as measured in Ref.\
\protect\onlinecite{pfl01} for ZrZn$_2$ is shown in Fig.\
\ref{pd}.  As noted by the authors of Ref.
\protect\onlinecite{pfl01}, one of the most intriguing and perhaps
surprising features of superconductivity in ZrZn$_2$ (as well as
that occurring in UGe$_2$) is that it occurs only in the
ferromagnetic state. The reason for the surprise is that
previous theoretical work had not anticipated that
superconductivity could occur in the ferromagnetic phase, unless
at the very least it was also stable in the paramagnetic phase.
The possibility that superconductivity might appear only in the
ferromagnetic phase does not seem to have been considered before
the recent experimental discoveries. For example, a very early
article \protect\cite{gin57} had noted that the presence of the
large internal magnetic induction in a ferromagnet would suppress
superconductivity.  Also, another early theoretical article
\protect\cite{fay80}, which demonstrated how spin fluctuations can
give rise to $p$-wave superconductivity, found that the
superconductivity occurs  in both the ferromagnetic and
paramagnetic phases  close to the ferromagnetic quantum critical
point (see  Fig.\ \ref{pd2}).  Other examples which find
superconductivity on the paramagnetic side of a ferromagnetic
quantum critical point include Refs.\
\protect\onlinecite{maz97,mon99,bla99,rou00}. Very recently it has
been argued \protect\cite{kir01} that the critical temperature for
spin-triplet $p$-wave superconductivity mediated by spin
fluctuations is generically much higher in the Heisenberg
ferromagnetic phase than in the paramagnetic phase, due to the
coupling of the magnons to the longitudinal spin susceptibility,
and this result is qualitatively in agreement with the
superconducting phase diagram for UGe$_2$
(see Fig.\ \ref{pd2}). Another line of argument
\protect\cite{mac01} is that the pairing symmetry realized in
UGe$_2$ must be a nonunitary spin-triplet pairing similar to that
realized \protect\cite{amb73} in the A$_1$-phase of superfluid
$^3$He  because such states are free from the Pauli limit and can
survive in a huge internal magnetic field. In addition, the
superconducting order-parameter symmetry in the ferromagnetic
phase of UGe$_2$ has been studied \protect\cite{fom01} in terms of
the magnetic point group symmetry of the ferromagnetic phase. The
ideas introduced below have some overlap with these latter
\protect\cite{mac01,amb73,fom01} ideas. Finally, we note an
article that has shown theoretically that coexisting
superconductivity and ferromagnetism can occur for the case where
the same band electrons produce both phenomena
\protect\cite{kar01}.

%old figure placement

This article describes a phenomenological model that gives a good
description of superconductivity in ferromagnetic ZrZn$_2$
(although not in ferromagnetic UGe$_2$).  In particular, the model
gives a natural explanation of the fact that superconductivity
occurs in the ferromagnetic but not in the paramagnetic phase. The
basic idea is that in the superconducting state the Cooper pairs can
have magnetic moments  --- see Ref.\ \protect\onlinecite{min99}.
In the presence of a ferromagnetic magnetization density in the
sample the magnetic moments of the Cooper pairs can interact with
this ferromagnetic magnetization density via an interaction having
the form of an exchange interaction. The Cooper pair magnetization
density chooses a direction that makes this ``exchange'' energy
negative, and this is the mechanism that makes the superconducting
state more stable in the ferromagnetic state than in the
paramagnetic state.  As will shown below, in order to give rise to
superconductivity in the ferromagnetic state but not in the
paramagnetic state, the exchange coupling just described must be
greater than a certain critical value.

The ferromagnetic state will be modelled using the Landau free
energy \protect\cite{lan60}
\begin{equation}
    F_f = \alpha^\prime_f [T - T_f(P)] M^2
   + \frac{1}{2} \beta_f M^4
     \label{FF}.
\end{equation}
Here the ferromagnetic transition temperature $T_f$ is assumed to
depend on the pressure P.  Expanding $T_f(P)$ in a Taylor series
about the point $P_c$ at which it goes to zero, and keeping only
the first nonvanishing term, yields $T_f(P) = T_f^\prime(P_c -
P)$. This linear dependence of $T_f$ on $P$ agrees well with the
experimentally measured pressure dependence
for ZrZn$_2$ \protect\cite{pfl01}, shown in Fig.\ \ref{pd}. From
Eq.\ (\ref{FF}) one finds $M = (\alpha_f^\prime /
\beta_f)^{1/2}(T_f (P) - T)^{1/2}$.

For cubic ferromagnets (such as ZrZn$_2$) the only two
possibilities  for the easy direction of the ferromagnetic
magnetization density are the a [100] or a [111]
direction \protect\cite{lan60}. Although in the absence of
ferromagnetism, the C15 Laves phase structure of ZrZn$_2$ has
cubic ${\bf O}_h$ (m3m) point group symmetry, the point group
symmetry in the presence  of ferromagnetism is reduced to the
magnetic point group ${\bf D}_{4h} ({\bf C}_{4h})$  $(4/m m^\prime
m^\prime)$ symmetry for the ferromagnetic magnetization density in
the [100] direction,
or ${\bf D}_{3d} ({\bf C}_{3i})$ $ (\overline{3}
m^\prime)$ for the ferromagnetic magnetization density in the
[111] direction. Since all of the irreducible representations of
${\bf C}_{4h}$ and ${\bf C}_{3i}$ are one dimensional, it is
expected that the transition to superconductivity in the presence
of ferromagnetism can be described by a one-component order
parameter.

It is of interest to investigate how this
one-component order parameter describing superconductivity in the
presence of ferromagnetism might be related to order parameters
appropriate to the description of superconductivity in cubic
ZrZn$_2$ in the absence of ferromagnetism. An advantage of treating the
paramagnetic, non-superconducting state as the reference state is that an
explicit dependence of the parameters describing the superconductivity on
$M$ is obtained (see below). Because of the large value of the exchange field
compared to the superconducting critical temperature, all spin-singlet
states of Cooper pairs are excluded. Thus, assuming spin-triplet pairing, 
consider
the representation  $F_{1u}$ of the group ${\bf O}_h$ for which the
order parameter is the three-component quantity
%$\mbox{\boldmath$\psi$} = (\psi_x, \psi_y, \psi_z)$
$\bm{\psi} = (\psi_x, \psi_y, \psi_z)$
whose components transform
under rotations like those of a three-dimensional
polar vector \cite{vol85} (the $F_{2u}$ representation gives the same model).
We use a strong spin-orbit coupling scheme, in which rotations
transform both spin and orbital degrees of freedom.
Also, the time-reversed state corresponding to  $\bm{\psi}$ is
$\bm{\psi}^R = (\psi_x^\ast , \psi_y^\ast,
\psi_z^\ast)$. Now define the vector product
\begin{equation}
     \bm{S} = i\bm{\psi}^*\times \bm{\psi}.
     \label{S}
\end{equation}
Because this quantity transforms like a magnetization density under the
operations of ${\bf O}_h$ and time reversal, it will be
interpreted  (to within a constant factor) as a magnetization density
associated with the
Cooper pairs. It should be noted that, at the
phenomenological level of this article,  in the strong
spin-orbit coupling scheme, the spin and orbital magnetization density of
Cooper pairs can not be distinguished.

Now consider the following terms of an expansion of the
Ginzburg-Landau (GL) free energy in powers of the components of the
order parameters $\bm{M}$ and  $\bm{\psi}$
\begin{equation}
F_{S,0} = \alpha \bm{\psi}^\ast \cdot \bm{\psi}
     - 4\pi J \bm{M} \cdot \bm{S}.
     \label{FS}
\end{equation}
Only the terms quadratic in the superconducting order parameter
and consistent with the cubic symmetry and time-reversal
invariance have been included here, since these are all that are
necessary (together with the gradient terms) to find the upper
critical field for superconductivity. Furthermore, terms up to
linear order in $\bm{M}$ have been included.  Note that the last
term in this equation has the form of an exchange interaction
between the ferromagnetic magnetization density and the
Cooper-pair magnetization density. If the exchange parameter $J$
is positive, the formation of a superconducting state in which the
Cooper-pair magnetization density is parallel to the ferromagnetic
magnetization density is favored. Also, if the ferromagnetic
magnetization density $\bm{ M}$ is rotated (by an applied magnetic
field) this exchange mechanism for stabilizing the
superconductivity is still applicable, and the orientation of the
Cooper-pair magnetization density $\bm{ S}$ will follow that of
the ferromagnetism. (If $J < 0$, an equivalent model is obtained
in which a Cooper-pair magnetization density antiparallel to the
ferromagnetic magnetization density is favored.) The free energy
of Eq.\ (\ref{FS}) is reminiscent of that employed in the
description of the A$_1$-phase of  $^3$He \protect\cite{amb73}.

Now call the direction of the nonzero ferromagnetic magnetization
density the $z$ direction (which, as noted above, can be either a
[100] or a [111] direction for a cubic ferromagnetic). In addition
to the exchange field coupled with the spin of electrons, the
magnetization creates an internal magnetic induction which interacts
with the electron charge. Thus, the superconductor should be
in the mixed state even in the absence of an external magnetic
field, and, in order to calculate the transition temperature, one
has to take into account the gradient terms in the GL free energy,
in addition to the uniform terms given by Eq.\ (\ref{FS}):
\begin{eqnarray}
F_S &=& F_{S,0}+K_1(D_i\psi_j)^*(D_i\psi_j)+K_2[(D_i\psi_i)^*(D_j\psi_j)
  \nonumber\\
  && +(D_i\psi_j)^*(D_j\psi_i)]+K_3(D_i\psi_i)^*(D_i\psi_i).
\label{FSdiag}
\end{eqnarray}
Here $\alpha=\alpha^\prime(T-T_0)$, and $T_0$ is the
superconducting transition temperature in the absence of
the exchange interaction of Cooper pairs with the ferromagnetic magnetization
(i.e. at $J=0$), which is assumed to be positive.
 The gradient part contains terms which are invariant under
rotations from the cubic group \cite{sig91}, with
$D_i=-i\hbar(\partial/\partial x_i)+(2|e|/c)A_i$,
and $\text{ curl} \bm{ A}=\bm{ B}$. The magnetic induction is given by
$\bm{ B}=\bm{H}_{ext}+4\pi\bm{M}$, where $\bm{H}_{ext}$ is
the external magnetic field directed along $z$, and
the magnetization density is
$\bm{M}=\bm{M}_0+(\mu-1)\bm{ H}_{ext}/(4\pi)$. The long cylinder
geometry, with the $z$ axis along the axis of the cylinder, has been
assumed.

Using a variational approach \cite{moo} to minimize the free energy
(\ref{FSdiag}), we calculate the superconducting critical temperature
as a function of the magnetization density and external field,
which takes the simple form
\begin{equation}
\label{Tc}
  T_c(M) = T_0+\frac{4\pi(J-J_c)}{\alpha'}M
\end{equation}
at $H_{ext}=0$. The quantity $J_c$ describes the suppression of
the critical temperature due to orbital effects. It takes
different values for $\bm{M}\parallel[100]$: $J_c=(|e|/\hbar
c)(2K_1+2K_2+K_3)$, and for $\bm{M}\parallel[111]$:
$J_c=(|e|/\hbar c)(2K_1+2K_2+2K_3/3)$. Another result of our
calculation is that the only component of the order parameter
which is non-zero at $T=T_c(M)-0$ is
$\psi_-=(\psi_x-i\psi_y)/\sqrt{2}$ (or $\psi_+=(\psi_x \pm
i\psi_y)/\sqrt{2}$ for $J < 0$). It is this quantity that
describes the formation of a superconducting state with its
Cooper-pair magnetization density parallel to the ferromagnetic
magnetization density.  Finally, note from Eq.\ ({\ref{Tc}) that,
in order for $T_c$ to be enhanced in the ferromagnetic phase
relative to its value $T_0$ in the paramagnetic phase, the
exchange parameter $J$ must be greater than $J_c$. In the
weak-coupling theory, $J$ is proportional to $N'(\epsilon_F)$, the
derivative of the single-particle density of states (DoS) at the
Fermi level \cite{min99}. The smallness of this quantity in
${^3}$He explains the narrow region of existence of the
A$_1$-phase. In the case of ZrZn$_2$, however, where the DoS is
extremely sharply peaked near the Fermi energy \cite{huang88},
$N'(\epsilon_F)$ could be very large, but  estimating $J$ in terms
of $N'(\epsilon_F)$ is probably too simplistic.

In order to confirm $\psi_-$ as a possible order parameter
describing the formation of superconductivity in the ferromagnetic
state, it should be checked that it transforms as a basis vector
of some irreducible representation of the magnetic symmetry group
of the ferromagnet \protect\cite{fom01}.  Suppose that the
ferromagnetic magnetization density is along the [100] direction.
Then the magnetic symmetry group is ${\bf D}_{4h}({\bf C}_{4h}) =
{\bf C}_{4h} + (RC_{2x}){\bf C}_{4h}.$ (Here R is the
time-reversal transformation.) In this case, $\psi_-$ transforms
like one of the complex irreducible representations ($^1$E or
$^2E$) of ${\bf C}_{4h}$. Furthermore, although there is no time
reversal operation in this magnetic group, the operator $RC_{2x}$
has the effect of replacing $\psi_-$ by its complex conjugate.
Hence $\psi_-$ is a possible order parameter.  A similar analysis
can be performed if the ferromagnetic magnetization density is
along the [111] direction, when the magnetic point group is ${\bf
D}_{3d}({\bf C}_{3i}) = {\bf C}_{3i} + (RC_{2x}){\bf C}_{3i}$.
Here too the order parameter transforms like one of the complex
representations ($^1$E or $^2E$) of the relevant point group
$({\bf C}_{3i})$.  It should be noted that this predicted symmetry
of the superconducting state can be verified by experimental
measurement (e.g. see \cite{sam02}).

The pressure dependence of the critical temperature $T_s$ of the
transition to the superconducting state can be found from Eq.\
(\ref{Tc}) to be given by the solution of
\begin{equation}
     T_s = T_0 + T^{*1/2}[T_f (P)-T_s ]^{1/2}
\label{Tsup}
\end{equation}
where
\begin{equation}
     T^* = \left(\frac{\alpha_f^\prime}{\beta_f}\right)
         \left(\frac{4\pi}{\alpha^\prime}\right)^2
         \left(J-J_c\right)^2.
     \label{T*}
\end{equation}

By assumption, the exchange enhancement results in a
superconducting transition temperature $T_s$ much greater than the
superconducting transition temperature $T_0$ in the paramagnetic
state.  Furthermore, except for $P$ very close to $P_c$, $T_s \ll
T_f(P)$.  Under these conditions the pressure dependence of $T_s$
is given by the formula
\begin{equation}
     T_s(P) = T_s(0)(1 - P/P_c)^{1/2}.
     \label{Ts}
\end{equation}
When $P$ gets very close to $P_c$ and $T_f(P)$ becomes very small
this equation is no longer valid.  In this extreme circumstance,
if one takes $T_0 = 0$ and $T_f(P) \ll T^\ast$, one finds
\begin{equation}
     T_s(P) = T_f(P)[1 - T_f(P)/T^\ast + ...]
\label{TsPc}
\end{equation}
which shows that, for $P$ very close to $P_c$ and $T_0 = 0$,
$T_s(P)$ approaches $T_f(P)$, and is never greater than $T_f(P)$.
Eq.\ (\ref{Ts}), together with the equation for the ferromagnetic
transition temperature
\begin{equation}
     T_f(P) = T_f(0)(1 - P/P_c)
     \label{Tf}
\end{equation}
arrived at above, have been used to plot the phase diagram of
Fig.\ (\ref{pd}), which shows a remarkable similarity to the
experimentally determined \protect\cite{pfl01} phase diagram for
ZrZn$_2$.

Finally, at a given pressure, we find a
temperature dependence of the upper critical field of
\begin{equation}
H_{c2}(T,P) = H_{c2}(0,P)[1 - T/T_s(P)] \label{hc2T}
\end{equation}
This is in reasonable agreement with the experimental result (see
Fig. 3 of Ref.\ \protect\onlinecite{pfl01}), which however has
somewhat more curvature than the linear temperature dependence
shown here.  The lack of curvature in the result of Eq.\
(\ref{hc2T}) results from the linear dependence of $M_z$ on
$H_{ext}$ in our relation $M_z = M_{z0} + (\mu -
1)H_{ext}/(4\pi)$.

We conclude that the proposed mechanism of stabilizing
superconductivity in a ferromagnet (by an exchange type of
interaction between the magnetization density of the Cooper pairs
and the ferromagnetic magnetization density) gives an excellent
qualitative description of the phase diagram determined
experimentally for ZrZn$_2$.  In particular, it explains in a
natural way the fact that the superconductivity occurs in the
ferromagnetic phase, but not in the paramagnetic phase.  For this
mechanism to work, the exchange interaction parameter must have a
magnitude larger than a certain critical value.  A further
experimental test of our model would be the determination of the
order parameter symmetry. (A prediction of our model is that the
order parameter transforms like one of the complex representations
of the relevant point group.)

It should be mentioned that the spin-fluctuation mechanism studied
in Ref. \cite{kir01} can provide an alternative explanation of
growing $T_s$ in the ferromagnetic state, given that the
magnetization in ZrZn$_2$ does not reach saturation.  To what
extent the fluctuation effects discussed in Ref. \cite{kir01} are
essential compared to the mean field interactions studied in this
article, is in our view still an open question, and their relative
contributions can be different in different materials. For
example, it seems that the phase diagram of UGe$_2$ (see Fig.
\ref{pd}) can be satisfactorily explained by the spin-fluctuation
theory, and the apparent absence of contributions from the
exchange interaction of our work could be explained by the
magnitude of the exchange parameter being less than its critical
value.

We acknowledge support from the Canadian Institute for Advanced
Research and from the Natural Sciences and Engineering Research
Council of Canada.

%\begin{references}

%\end{references}

\end{document}